\documentclass[12pt,a4paper]{article}
\makeatletter
\newcommand*{\indep}{%
  \mathbin{%
    \mathpalette{\@indep}{}%
  }%
}
\newcommand*{\nindep}{%
  \mathbin{
    \mathpalette{\@indep}{/}%
  }%
}
\newcommand*{\@indep}[2]{%
  \sbox0{$#1\perp\m@th$}
  \sbox2{$#1=$}
  \sbox4{$#1\vcenter{}$}
  \rlap{\copy0}
  \dimen@=\dimexpr\ht2-\ht4-.2pt\relax
  \kern\dimen@
  \ifx\\#2\\%
  \else
    \hbox to \wd2{\hss$#1#2\m@th$\hss}%
    \kern-\wd2 %
  \fi
  \kern\dimen@
  \copy0 
}
\makeatother
\usepackage{dsfont}
\usepackage{rotating}
\usepackage{float}
\usepackage{siunitx}
\usepackage{mathrsfs}  
\usepackage[latin1]{inputenc}
\usepackage{amsmath}
\usepackage{amssymb}
\usepackage{amsfonts}
\usepackage{graphicx}
\usepackage{color}
\usepackage[numbers]{natbib}
\usepackage{array} 
\usepackage{mathrsfs}  
\usepackage{url}  
\usepackage{natbib}
 \usepackage{mathabx}
\usepackage{pgf,tikz}
\usepackage{booktabs} 
\usepackage{colortbl} 
\usepackage{xcolor} 
\usepackage{xfrac}
\usepackage{makecell}
\usetikzlibrary{cd}
\newcommand{\blind}{1}
\begin{document}

\def\spacingset#1{\renewcommand{\baselinestretch}%
{#1}\small\normalsize} \spacingset{1}
\date{ }


\if1\blind
{
\title{\bf Using functional information for binary classifications} 
\author{Pablo Mart\'inez-Camblor$^{1,2}$\thanks{{\it Correspondence to:} Pablo Mart\'inez-Camblor. 7 Lebanon Street, Suite 309, Hinman Box 7261, Lebanon, NH 03751, USA. E-mail: {\color{blue} Pablo.Martinez-Camblor@hitchcock.org}}
\hspace{.2cm}\\
\small{$^1$Departments of Anesthesiology and Biomedical Data Science,}\\ \small{Geisel School of Medicine at Dartmouth, NH, USA} \\
\small{$^2$Faculty of Health Sciences, Universidad Autonoma de Chile, Chile}}
\maketitle
} \fi

\begin{abstract}
The adequate use of information measured in a continuous manner along a period of time represents a methodological challenge. In the last decades, most of traditional statistical procedures have been extended for accommodating these functional data. The binary classification problem, which aims to correctly identify units as positive or negative based on marker values, is not aside of this scenario. The crucial point for making binary classifications based on a marker is to establish an order in the marker values, which is not immediate when these values are presented as functions. Here, we argue that if the marker is related to the characteristic under study, a trajectory from a positive participant should be more similar to trajectories from the positive population than to those drawn from the negative. With this criterion, a classification procedure based on the distance between the involved functions is proposed. Besides, we propose a fully non-parametric estimator for this so-called probability-based criterion, PBC. We explore its asymptotic properties, and its finite-sample behavior from an extensive Monte Carlo study. The observed results suggest that the proposed methodology works adequately, and frequently better than its competitors, for a wide variety of situations when the sample size in both the training and the testing cohorts is adequate. The practical use of the proposal is illustrated from real-world dataset. As online supplementary material, the manuscript includes a document with further simulations and additional comments. An R function which wraps up the implemented routines is also provided.
\end{abstract}
\noindent%
{\it Keywords: Area under the curve; Binary classification; Functional data; Non-parametric model; ROC curve.}
\vfill

\spacingset{1.45} 
\section{Introduction}
Modern technology devices are able to generate and storage large quantities of information. Not infrequently, this information includes different variables of interest measured in a continuous-manner along a period of time. The adequate use of the resulting functional data represents a methodological challenge. In the last decades, most of traditional statistical procedures have been extended for accommodating this type of data. For an overview of the topic see, for instance, the monographs by \citet{ramsay06}, and Ferraty and Vieu \cite{ferraty06}. Theoretical and inferential aspects can be found in more recent books by \citet{horvath12}, and \citet{koloszka18}, among others. The binary classification problem, which aims to correctly identify units as positive or negative based on marker values, is not aside of this scenario. We recommend the excelent review of  \citet{wang24} for having an overview about several emerging and promising areas of functional data classification.

The receiver operating characteristic, ROC, curve \cite{lusted71} is a graphical tool routinely used for representing the performance of continuous markers in terms of the sensitivities (probability that a positive participant is correctly classified as positive) and specificities (probability that a negative participant is correctly classified as negative) associated with {\it all} potential thresholds. Besides, the area under the curve, AUC, is a popular index for measuring the overall discrimination capacity of the marker through a single number \cite{hanley82}. Both theoretical and practical aspects of ROC curves and AUCs have been broadly treated in the specialized literature (see, for instance, the monographs of \citet{zhou02}, \citet{pepe03}, or \citet{nakas23}, among others). 

The crucial point for making binary classifications based on a marker is to establish an order in the marker values. When the considered marker incorporates multivariate information, usual practice is searching for linear combinations maximizing the resulting AUC (see \citet{sonia20} for an overview of existing methods). \citet{estevez21} considered markers living in an infinite-dimensional space (functions). In their paper, the sample of functions is ranked via a projection over a particular subset of the functional space adequately indexed (ordered). This approach identifies difference in the location parameter, but it is not adequate when main difference between the structure of the functions drawn from the positive and the negative groups is in their shape or covariance structure \cite{bianco24}. \citet{jang23} reduced each function to a single value. They considered a number of important and simple pharmacokinetic features such as the minimum or the maximum values, but also integral-type summaries such as the average velocity or the average acceleration. \citet{bianco24} extended the idea of finding linear combinations for maximizing the AUC of the resulting punctuation to the functional data framework. They also considered quadratic transformations which allow to deal with cases where the involved covariance structures are different. Besides, a number of extensions for regression models including functional predictors have been proposed in the specialized literature. These proposals include binary dependent variables, which are strongly related to the binary classification problem, and that can be used with this goal (see, for instance, \citet{escabias07} \cite{escabias14}, among others). Here, we argue that if the marker is related to the characteristic under study, a trajectory from a positive participant should be more similar to trajectories from the positive population than to those drawn from the negative. Based on that, we transform each trajectory in a probability of being positive (against of being negative), and propose to use the resulting probabilities for allocating the participants within the positive or within the negative group. We call this criterion PBC (probability-based classification).

Rest of the paper is organized as follows. In Section 2, we present an overview of how functional data can be used in the binary classification problem, introduce the proposed criterion, and derive a fully non-parametric estimator. In Section 3, we consider asymptotic properties, discuss some theoretical limitations of our proposal, and explore the use of different inferential approaches. Section 4 shows the results of the Monte Carlo simulations conducted in order to study the finite-sample behavior of the PBC criterion. In Section 5, we explore in deep the problem of identifying breast cancer patients at risk of having cardiotoxicity therapy-related cardiac dysfunction (CTRCD) through the functional transformation of the results obtained by Tissue Doppler Imaging (TDI) testing. This dataset is publicly available at {\color{blue}{\url{https://figshare.com/articles/dataset/BC_cardiotox_A_cardiotoxicity_dataset_for_breast_cancer_patients/22650748/4}}}, and fully explained in \citet{pineiro23}. Same dataset has been previously considered by \citet{bianco24} for illustrating their proposal. Our main conclusions are presented in Section 6. Technical details and other complements are provided in the Appendix and in the online document Supplementary Material, respectively.

\section{Classification based on functions}
We consider here the use of functions for classifying subjects (units, participants, patients) as positive or negative. Let $\mathscr F$ be a family of functions which constitutes a separable Hilbert space. Particularly, we consider that each $f\in \mathscr F$ is a function such that $f:T\rightarrow\mathbb R$, with $\int_T f(t)^2dt<\infty$ for some compact $T\subset\mathbb R$. Let $\mathscr F_N$ and 
$\mathscr F_P$ be subsets of $\mathscr F$ representing the trajectories drawn from the negative ($Y=0$), and from the positive ($Y=1$) populations, respectively. Notice that, in order to have some discriminatory ability, the functions within the sub-families $\mathscr F_N$ and $\mathscr F_P$ should have different behavior.

Following the overall {\it classification subsets} approach introduced by \citet{camblor19}, classifying participants based on the information provided by a functional marker implies to define a criterion for being positive (analogously, we could define a criterion for being negative), ${\cal C}_p$ ($0\leq p\leq 1$), such that
$$\mathbb P(f\in {\cal C}_p| Y=0)=p.$$
That is, following this classification criterion, the specificity of the (functional) marker is $1-p$. Therefore, the sensitivity will be determined by
$$\mathbb P(f\in {\cal C}_p| Y=1).$$
Once defined a family of classification criteria, $\{ {\cal C}_p\}_{\{0\leq p\leq 1\}}$, the ROC curve associated with these classification rules would be determined by the pairs 
$$\left\{\mathbb P(f\in {\cal C}_p| Y=0),\, \mathbb P(f\in {\cal C}_p| Y=1)\right\}=\left\{p,\, \mathbb P(f\in {\cal C}_p| Y=1)\right\},$$
and therefore, the ROC curve at point $p$ ($0\leq p\leq 1$) will be 
\begin{equation}
{\cal R}(p)=\mathbb P(f\in {\cal C}_p| Y=1).
\end{equation}
Given the nature of the problem, the immediate temptation (that would avoid ordering the functions) would be to define a region ${\cal I}_p\subset T\times\mathbb R$, and consider 
$f\in {\cal C}_p:=\{f\in \mathscr F: (t,f(t))\in {\cal I}_p,\, \forall t\in T\}$ (similar, although alternative, formulations of these regions could be easily defined). Following \citet{camblor17x}, for each $p\in [0,1]$, we will search for a ${\cal C}_p$ such that $\mathbb P(f\in {\cal C}_p| Y=0)=p$, and which maximizes $\mathbb P(f\in {\cal C}_p| Y=1)$. Unfortunately, in this context, this approach has two major wrinkles. $i)$ The estimation of the regions based on finite samples would be strongly overfitted, and the extrapolation of the results would be probably complex. $ii)$ The criterion would miss the longitudinal nature of the information contained in the trajectories, which could mislead the problem for particular curve configurations (see, for instance, Model II, in Section 4).

Alternatively, as it is reported in \citet{bianco24}, we can define an operator, $\Upsilon:\, {\mathscr F}\rightarrow {\mathbb R}$, which transforms the elements of $\mathscr F$ (functions) in real values. Then, we can directly use these real values for classifying. As we have mentioned in the Introduction, examples of $\Upsilon(\cdot)$ considered in the existing literature are the minimum value, the maximum value, or the integral, among many others.

\subsection{Probability-Based Classification}

We argue that functions drawn from $\mathscr F_P$ ($\mathscr F_N$) would be closer to functions from $\mathscr F_P$ ($\mathscr F_N$) than to those functions from $\mathscr F_N$ ($\mathscr F_P$). Therefore, for each $f\in \mathscr F$, we define the random variable
\begin{equation}
X_f:= \left\{d_f(g):= \int_T (f(t)-g(t))^2dt,\, g\in {\mathscr F}\right\}.
\end{equation}
And, directly
\begin{equation}
X_{f,k}:= \left\{d_f(g):= \int_T (f(t)-g(t))^2dt,\, g\in {\mathscr F}_k\right\},\quad k\in \{N,P\}.
\end{equation}
Our assumption implies that, when $f\in \mathscr F_k$, the smallest values of $X_f$ would correspond to $X_{f,k}$ ($k\in \{N,P\}$), and therefore, for an adequate transformation, $h_f(\cdot)$, the random variable
\begin{equation}
{\cal P}:= \left\{p_f:= \mathbb P(h_f(X_{f,P})< h_f(X_{f,N})), f\in \mathscr F\right\},
\end{equation}
takes larger values for $f\in \mathscr F_P$ than for $f\in \mathscr F_N$. We propose to classify the participants using the criteria
\begin{equation}
{\cal C}_p:=  \left\{f\in {\mathscr F}:\,  p_f > F_0(1-p) \right\},\quad 0\leq p \leq 1,
\end{equation}
where $F_0(\cdot)={\mathbb P}({\cal P}\leq \cdot| Y=0)$, that is, the cumulative distribution function (CDF) of ${\cal P}$ in the negative population. Therefore, for $0\leq p \leq 1$, the ROC curve associated with the PBC criterion is defined by 
\begin{equation}
{\cal R}(p)= 1 - F_1(F_0^{-1}(1-p)),
\end{equation}
with $F_1(\cdot)={\mathbb P}({\cal P}\leq \cdot| Y=1)$, CDF of ${\cal P}$ in the positive population.\par
\phantom b\par
\medskip\noindent
{\bf Remark 1.} When the main difference between curves from the positive and the negative populations are in the location parameter, $\mathbb P(X_{f,P}<X_{f,N})$ would be larger for $f \in X_{f,P}$ than for $f \in X_{f,N}$. However, we want that our procedure considers situations in which a positive function could be close only to a particular subgroup of positive functions. The role of $h_f(\cdot)$ is dealing with potential mixture behaviors of $\mathscr F_P$ ($\mathscr F_N$), or large difference in the variance structure of  $\mathscr F_N$ and  $\mathscr F_P$. In these cases, to be close of a subset of positive functions but far from others would be in the line of our argument for being classified as positive. In this sense, $h_f(\cdot)$ helps to ponder the group with the smallest differences. The gROC curve  \cite{camblor17x} allows a fully non-parametric estimation of $h_f(\cdot)$ \cite{camblor22, camblor25}, always considering that smaller distances with the positive functions have to be associated with higher probabilities of being positive.\par

\subsection{Empirical estimator for the PBC}

Let $\{f_1,\cdots , f_{n_0}, f_{n_0 +1},\cdots , f_{n_0 + n_1}\}$ be a sample of $n$ ($=n_0+n_1$) independent random functions drawn from $\mathscr F$, where the first $n_0$ units are from $\mathscr F_N$, and the remaining $n_1$ from $\mathscr F_P$. For each function, $f_i$ ($1\leq i \leq n$) we compute the random variable 
$$\hat X^n_{f_i}:=\left\{\hat d_{f_i,n}(f_j):= \int_T (f_i(t)-f_j(t))^2dt,\, i\neq j\in \{1, \cdots, n\}\right\}.$$
We consider immediate the definitions of $\hat X^{n_0}_{f_i,N}$  and $\hat X^{n_1}_{f_i,P}$. Then, once estimated the transformation, $h_f(\cdot)$, we use the well-known Mann-Whitney estimator for approximating the probability ${\mathbb P}(\hat h_f(X_{f,N})< \hat h_f(X_{f,P}))$. Therefore, for $1\leq i\leq n_0$, we define
$$\hat p_{f_i,n}:= \frac{1}{(n_0-1)\cdot n_1}\sum_{\substack{ j=1\\ j\neq i}}^{n_0}\sum_{k=n_0+1}^{n_1+n_0}{\mathds 1}\{\hat h_{f_i}(\hat d_{f_i,n}(f_k))<\hat h_{f_i}(\hat d_{f_i,n}(f_j))\},$$
and, for $n_0< i\leq n_0+n_1$,
$$\hat p_{f_i,n}:=  \frac{1}{n_0\cdot (n_1-1)}\sum_{j=1}^{n_0}\sum_{\substack{k=n_0+1\\ k\neq i}}^{n_1+n_0}{\mathds 1}\{\hat h_{f_i}(\hat d_{f_i,n}(f_k))<\hat h_{f_i}(\hat d_{f_i,n}(f_j))\}.$$
where ${\mathds 1}\{A\}$ stands for the standard indicator function (takes value 1 if $A$ is true, and 0 otherwise). Finally, we have 
$$\hat{\cal P}_{n}:= \left\{\hat p_{f_i,n},\quad i\in \{1, \cdots , n_0+n_1\}\right\}.$$
The proposed estimator for the target ROC curve is,
\begin{equation}
\hat {\cal R}_{n,\bullet}(p)= 1 - \hat F_{n_1,n}(\hat F^{-1}_{n_0,n}(1-p)), \quad p\in [0,1],
\end{equation}
where $\hat F^{-1}_{n_0,n}(\cdot)=\inf\{s: \hat F_{n_0,n}(s)> \cdot \}$, with $\hat F_{n_k,n}(\cdot)$ the empirical cumulative distribution function (eCDF) for $F_{n_k}(\cdot)={\mathbb P}({\cal P}_n\leq \cdot| Y=k)$\, ($k\in\{0,1\}$).

Notice that, in $\hat F^{-1}_{n_k,n}(\cdot)$ ($k\in \{0,1\}$), the first subscript, $n_k$, refers to the 'sample' used for estimating the CDF, while the second, $n$, refers to the number of distances used for computing the $n_k$-values (actually, $n-1$, in this case). Similar criterion is used for labeling the ROC curve estimator. In $\hat {\cal R}_{n, m}(\cdot)$, $n$ refers to the sample used for computing the distances, while $m$ refers to the 'new' functions being classified. The estimator which uses the same set of functions for both tasks is labeled by $\hat {\cal R}_{n,\bullet}(c\dot)$. The set of functions used as reference for computing the distance of a given 'new' function will be called {\it system} or, for highlighting its random nature, {\it sample-system} (${\cal R}_{n}(p)=1 - F_{n_1}(\hat F^{-1}_{n_0}(1-p))$, $p\in [0,1]$, refers to the real ROC curve based on the sample-system '$n$').\par 

\phantom b\par
\medskip\noindent
{\bf Remark 2.} In practice, for applying the procedure to a 'new function' (patient) we need to have the {\it system} fed with a set of functions from which we are first computing the distances to the positive and to the negative populations, and then, the  probability associated to this 'new function'. This probability will be used for taking decisions. In this sense, with a machine learning perspective, the {\it system} could be fed with new and more specific information.

\section{Asymptotic properties}
Next result guarantees the uniform consistency of the non-parametric estimator proposed in Subsection 2.2. \par
\medskip
\noindent {\bf Theorem 1.} Let $\{f_1,\cdots , f_{n_0}, f_{n_0 +1},\cdots , f_{n_0 + n_1}\}$ be a random sample of $n$ functions drawn from a separable Hilbert space $\mathscr F$, where the first $n_0$ units are from $\mathscr F_N$ and the 
remaining $n_1$ from $\mathscr F_P$. If $n_1/n_0=\lambda^2_n\rightarrow\, \lambda^2<\infty$, and the related functions $\hat h_{f_i}(\cdot)$ ($1\leq i\leq n$) are consistently estimated. Then, if ${\cal R}(\cdot)$ has two continuous and bounded derivatives,
\begin{equation}
\sup_{p\in [0,1]} |\hat {\cal R}_{n,\bullet}(p) - {\cal R}(p)|\longrightarrow_n\, 0\quad a.s.\, \text{(almost surely).}
\end{equation}
\hfill $\Box$

We provide a proof of this result in the Appendix.

The asymptotic normality of the estimator is far from being immediate. Notice that the $n\times n$ matrix containing the distances between the functions is symmetric (with zeros in the main diagonal). The pairs $(\hat X^n_{f_i},\, \hat X^n_{f_j})$ ($1\leq i\neq j\leq n$) are not independent, and the relationships between these pairs are inherited by their associated probabilities. Therefore, standard asymptotic results regarding ROC curve and AUC are not directly applied to the PBC estimator. 

Fortunately, standard training-testing approach would break the dependency between the probabilities. We can use the testing cohort for constructing adequate confidence intervals on both the ROC curve and the AUC values. Based on that, we {\it i)} randomly select a portion of the sample for constructing the {\it sample-system}, and {\it ii)} compute ROC curve and AUC values using the remaining portion of the sample. The resulting ROC curve estimator, $\hat {\cal R}_{ns,nc}(\cdot)$ ($ns+nc=n$) satisfies  the following result.\par
\medskip
\noindent {\bf Theorem 2.} Under Theorem's 1 assumptions, let $\hat {\cal R}_{ns,nc}(\cdot)$ be the ROC curve estimator based on the training-testing algorithm described above. Then, if ${\cal R}_{ns}(\cdot)$ has a continuous and bounded derivative, $r_{ns}(\cdot)$, there exists a probability space on which one can define two independent sequences of Brownian bridges, ${\mathscr B}^{(nc_0)}_0\{p\}_{\{0\leq p\leq 1\}}$ and ${\mathscr B}^{(nc_1)}_1\{p\}_{\{0\leq p\leq 1\}}$, such that
\begin{equation}
\sqrt{nc_1}\cdot [\hat {\cal R}_{ns,nc}(p) - {\cal R}_{ns}(p)]= {\mathscr B}^{(nc_1)}_1\{1-{\cal R}_{ns}(p)\} + \lambda\cdot r_{ns}(p)\cdot {\mathscr B}^{(nc_0)}_0\{1-p\} + o(1)\quad a.s.
\end{equation}
where $nc_0$ and $nc_1$ are the number of negative and positive curves included in the testing cohort, respectively. 
\hfill $\Box$\par

Notice that, since the functions are estimated outside the considered sample, proof of this Theorem is directly derived from \citet{hsieh96}.

A direct consequence of Theorem's 2 is that, if $\hat{\cal A}_{ns,nc}=\int_0^1 \hat {\cal R}_{ns,nc}(p)dp$ and ${\cal A}_{ns}=\int_0^1 {\cal R}_{ns}(p)dp$, then we can immediately derive the following asymptotic convergence.\par
\medskip
\noindent {\bf Corollary 1.} Under the Theorem's 2 assumptions, we have that
\begin{equation}
\label{auc}
\sqrt{nc_1}\cdot [\hat {\cal A}_{ns,nc} - {\cal A}_{ns}]\stackrel{\cal L}{\longrightarrow}_{nc_1}\, {\mathscr N}(0, \sigma_\lambda),
\end{equation}
where $\sigma^2_\lambda=\| F_{ns_1}\|_{F_{ns_0}} + \lambda^2\cdot \| F_{ns_0}\|_{F_{ns_1}}$, with $ns_0$ and $ns_1$ the number of curves from the negative and the positive populations included in the training cohort, respectively, and $\| F\|_{G}=\int F^2(x)dG(x) - \left(\int F(x)dG(x)\right)^2$.
\hfill $\Box$\par
\phantom b\par
\medskip\noindent
{\bf Remark 3.} Logically, the distance between ${\cal R}_{ns}(\cdot)$, based on a sample-system with $ns$ functions, and the real ${\cal R}(\cdot)$, based on the real-system, depends on the number and representativity of the $ns$ functions included in ${\cal R}_{ns}(\cdot)$. In practice, none of them can be computed, and only the sample-system can be used for doing real classifications.

\section{Numerical simulations}
We consider here some Monte Carlo simulations for exploring the finite-sample behavior of the PBC criterion described previously. We randomly select 1/3 of the sample for constructing the model ({\it sample-system}) and use the remaining 2/3 of the sample for estimating ROC curves and AUCs. Since, beyond the quality of the estimator, the different criteria have different underlying realities, we will report the areas under the curve obtained in different models for the proposed estimator (PBC), the minimum (Min.), the maximum (Max.), and the integral (Int.) criteria considered in \citet{jang23}. Besides, we consider linear (Lin.) and quadratic (Qua.) approximations based on the functional logistic regression models implemented by \citet{escabias22} in the \url{R} \url{package} \url{logitFD} (details and used R code are provided as supplement). In general, we study four main functional scenarios, and then, we add different {\it types of noise} for obtaining the final trajectories. We run 1000 Monte Carlo simulations with four different sample size configurations ($(n_0, n_1)=(50,50),\, (50,100),\ (100,100),\, (200,100)$). Next, we briefly describe the models.\par
\phantom b\par
\noindent {\bf Model I.} We study raws based on the functions $g(t)=\sin(\pi\cdot t)$, and $f(t)=(7/5)\cdot\sin(\pi\cdot t)$, with $t\in [-1,1]$. In Model I-a, curves from both the negative and the positive samples were drawn from $g(t) + \epsilon_A(t)$, where $\epsilon_A(t)$ has a scaled Brownian motion structure with $\sigma(t)=1/200$ (real AUC is 0.5). In Model I-b, the curves from the negative population were also drawn from $g(t) + \epsilon_A(t)$, but curves from the positive population were from $f(t) + \epsilon_A(t)$. In Model-c, we introduced some heterogeneity in the variance structure; negative curves were again generated from $g(t) + \epsilon_A(t)$, but positive curves were from $f(t) + \epsilon_B(t)$, where $\epsilon_B(t)$ has a scaled Brownian motion structure with $\sigma(t)=1/100$. Finally, Model-d included asymmetric noises. Negative curves were drawn from $g(t) + \tau_A(t)$, and positive curves from $f(t) + \tau_A(t)$, where $\tau_A(t)$ is a centered Exponential variogram with parameter 200.\par
\phantom b\par
\noindent {\bf Model II.} We explore now the behavior of the different criteria on 'mixture functions'. In the Model II-a, for $t\in [-1,1]$, curves from the negative group are $g(t)=b\cdot t^2 \cdot {\mathds 1}\{t\leq 0\} - b\cdot t^2\cdot {\mathds 1}\{t>0\}$, and from the positive group $f(t)=a\cdot t^2$, where $a$ and $b$ are independently generated from two random variables with CDF $(1/2)\cdot \Phi_{-2,1/4}(\cdot) +(1/2)\cdot \Phi_{2,1/4}(\cdot)$, with $\Phi_{\mu,\sigma}(\cdot)$ the CDF of a normal random variable with mean $\mu$ and standard deviation $\sigma$. Then, we follow the same previous structure. In Model II-b, the curves are $g(t) + \epsilon_A(t)$ (negative), and $f(t) + \epsilon_A(t)$ (positive), with $\epsilon_A(t)$ scaled Brownian motions ($\sigma(t)=1/200$). For Model II-c, we consider the same functions for the negative group, but we run functions from $f(t) + \epsilon_B(t)$ in the positive, with $\epsilon_B(t)$ also scaled Brownian motions ($\sigma(t)=1/100$). Finally, in Model II-d, $g(t) + \tau_A(t)$, and $f(t) + \tau_A(t)$; $\tau_A(t)$ centered Exponential variograms with parameter 200.\par
\phantom b\par
\noindent {\bf Model III.} In Model III, the functions are the probability densities of a normal distribution with mean $\mu$ and standard deviation $\sigma$, $\varphi_{\mu,\sigma}(t)$ ($t\in [-1,1]$). In the Model III-a, for each curve from the negative group, the values of $\mu$ and $\sigma$ were randomly generated from $\mathscr N(-0.15, 0.1)$, and the absolute value of $\mathscr N(0.5,0.2)$, respectively ($\mathscr N(\mu,\sigma)$ normal variable with mean $\mu$ and variance $\sigma^2$). For the positive curves, the values of $\mu$ and $\sigma$ were generated from $\mathscr N(0.15, 0.1)$, and the absolute value of $\mathscr N(0.5,0.2)$, respectively. Then, we followed the previous structure. In Model III-b, again, we added scaled Brownian motions, $\epsilon_A(t)$, to all the curves, with $\sigma(t)=1/200$. In Model III-c, the negative functions follow the previous structure, but in the positive curves, we added scaled Brownian motions, $\epsilon_B(t)$, with $\sigma(t)=1/100$. In Model III-d, we added to both the negative and the positive trajectories centered Exponential variograms with parameter 200.\par
\phantom b\par
\noindent {\bf Model IV.} Finally, considered trajectories differ only at the end of the domain. We followed the same structure than in Model-I. In Model IV-a, both the negative and the positive functions were drawn from $g(t) + \epsilon_A(t)$, where $g(t)=-(1/2)\cdot t^3/(t-2)^2$, $\epsilon_A(t)$ has the same previous definition (real AUC is 0.5). Similarly, in Model IV-b, negative trajectories were again from $g(t) + \epsilon_A(t)$, while positive were from $f(t) + \epsilon_A(t)$, with $f(t)= (1/2)\cdot t^3/(t-2)^2$. In Model IV-c, the curves were drawn from $g(t) + \epsilon_A(t)$ (negative), and $f(t) + \epsilon_B(t)$ (positive), following consistent definitions. In Model IV-d, we incorporated the centered Exponential variogram and have the functions $g(t) + \tau_A(t)$, and $f(t) + \tau_A(t)$ for the negative and the positive populations, respectively. \par
\phantom b\par
\begin{figure}
\centerline{\includegraphics[width=17cm]{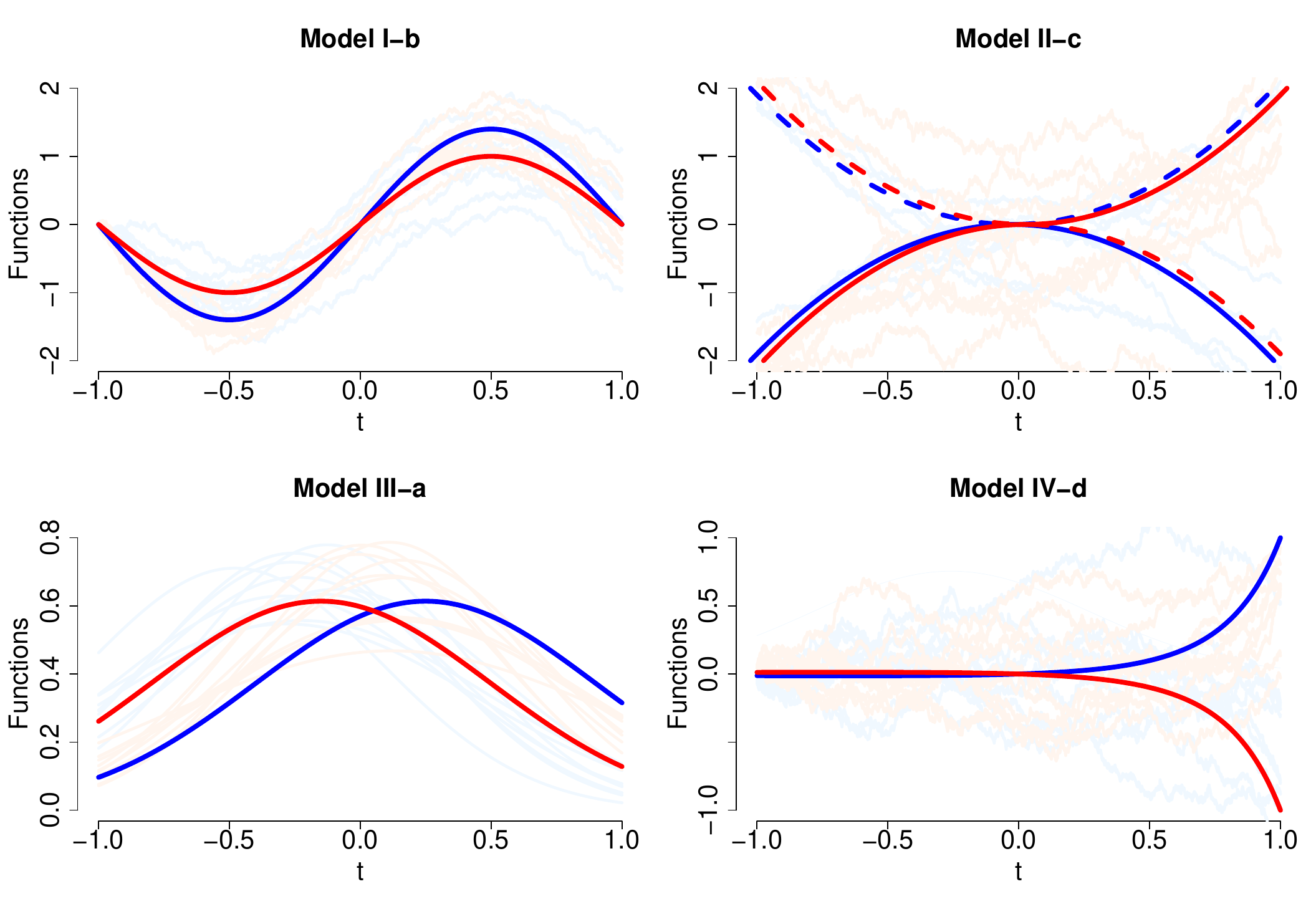}}
\caption {{\bf Models.} Average shape of the functions considered in the models I, II, III, and IV, with some random trajectories (shadow) from some of the particular models. \label{models}}
\end{figure}

Figure \ref{models} depicts the overall behavior of the four models and, in the shadow, some random trajectories as example. Table S1 (included in the online supplement material) contains the ROC curves and the AUCs for the 16 models studied based on a sample size of $(n_0, n_1)=(2500,2500)$ (in the PBC case, we run two independent samples with this size, one for training, one for testing). We will consider these values as the reality of reference ({\it real-system}).

\phantom b\par

\begin{figure}
\centerline{\includegraphics[width=15.5cm]{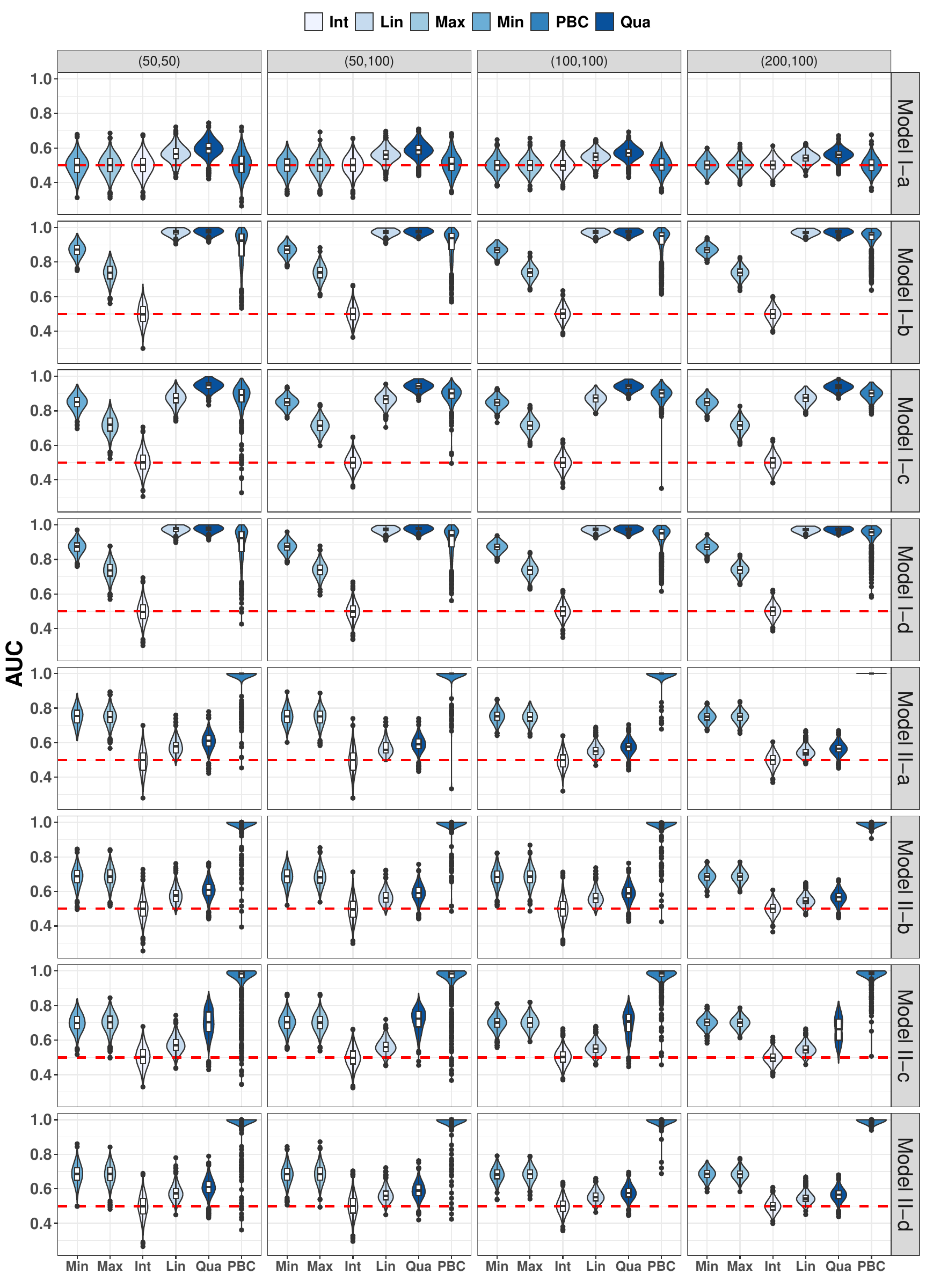}}
\caption {{\bf Models I.} Violin plots for the AUCs observed in 2,000 Monte Carlo simulations for the models type I, and II, with different sample sizes.\label{mc_01}}
\end{figure}

\begin{figure}
\centerline{\includegraphics[width=15.5cm]{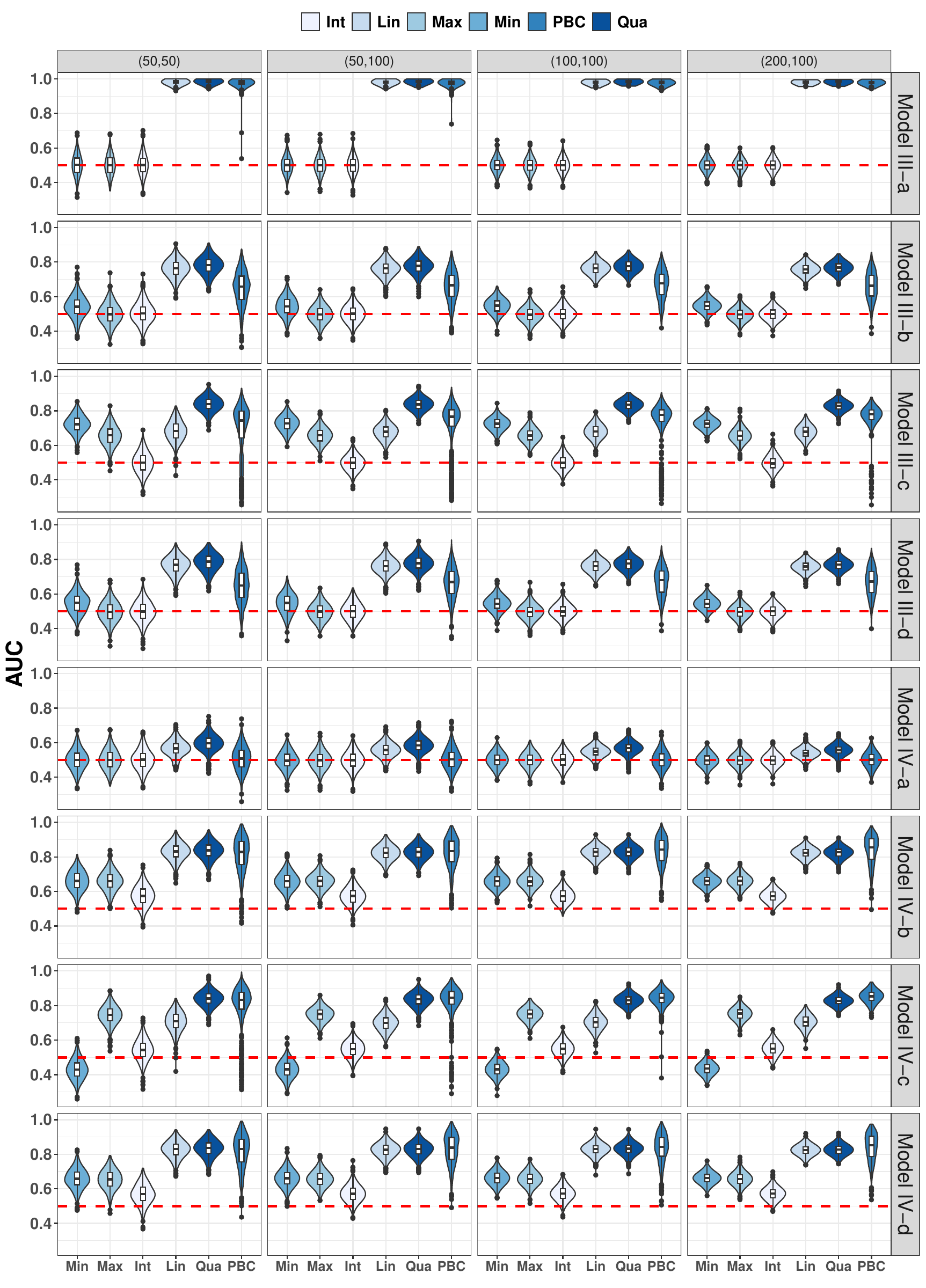}}
\caption {{\bf Models II.} Violin plots for the AUCs observed in 2,000 Monte Carlo simulations for the models type III, and IV, with different sample sizes.\label{mc_02}}
\end{figure}

Figures \ref{mc_01} and \ref{mc_02} show the violin plots for the AUCs obtained in our simulations. Models I-a, and IV-a emulate a situation in which the marker does not provide any information (AUC=0.5). We can see that Min., Max., Int., and PBC behave adequately and report average AUCs around 0.5. However, Lin. and Qua. show a clear overfitting problem, diluted but not fully removed when the sample size increases. In Models I-b, I-c, and I-d, Int., Max., and Min. had any, moderate, and good classification capacity, respectively. This capacity is not affected by the type of noise added. Lin., and Qua. showed strong classification (almost perfect) accuracies. Lin. was more affected than Qua. by the difference in the variance structure of the trajectories (Model I-c). PBC reached, in general, very good results, however, and not surprisingly, it shows strongly asymmetric distributions; the sample-system used for training can fail when the number of trajectories used is excessively low. The probability of randomly selecting a poor sample-system in a context of good classifications capacity dilutes clearly when the sample size increases. Despite that, in these models, the overall classification capacity competes with Lin. and Qua. In Model II, Lin. and Qua. provided poor results, slightly better than those obtained by Int., which is not related with the outcome. Highlight that despite Qua. takes some advantage of the different noise structure (Model II-c), its results still poor. Besides, Min. and Max. reported moderate results, not really affected by the structure of the noise. In these models, PBC is the clear winner. However, even in this very favorable situation, we can observe some outlier caused by small size in the sample-system. The problem disappears in Model II-(a,b,d) for sample size (200,100) (sample-size system based on 67 negative, and 33 positive trajectories). In general, in the Model III, Min., Max., and Int. were unable to distinguish between the two populations, although Min. and Max. have some benefits from the difference in the variance structure considered in Model III-c. Lin., Qua. reached an almost perfect classification in Model II-a, and good and similar results in configurations -b, -c, and -d. Again, highlight that Lin. suffers in scenarios with different variance structure (e.g. Model III-c). PBC shows a behavior dominated by high variability and the presence of outliners. It obtained similar results to Lin. and Qua. in -a, and even better than Lin. in -c. In the scenarios -b and -d, its results were disperse, with AUCs values ranging from 0.35 to 0.90 (mean of 0.65$\pm$0.1) for $(n_0,n_1)=$(50,50), and worse than those obtained by Lin. and Qua. Notice that, in the real-system, results of Lin., Qua., and PBC methods were similar (Table S1). Finally, in the Model IV, Int. was mostly unrelated with the outcome. Min., and Max. got moderate-low accuracies (better Max. in Model IV-c). Lin. and Qua. behaved good, better Qua. in Model IV-c. The performance of PBC was considerable good. Again, we have to highlight the great variability, although its results were, in general, slightly better than those obtained by Qua.

\phantom b\par

\phantom b\par

We also study the ability of the approximation provided in (\ref{auc}) for constructing accurate confidence intervals. For each previous Monte Carlo iteration, we compute a 95\% confidence interval, and we check whether the AUC derived from a fixed independent sample of size $(n_0, n_1)=(2500, 2500)$ (which will be considered as the real behavior of the sample-system, ${\cal R}_{ns}(\cdot)$) is within this interval. Table \ref{coverage} reports the observed coverage percentages of these 95\% confidence intervals, and their average lengths. Besides, it also includes the percentage of times that these intervals contain the {\it real-system} AUC based on the previously reported real-system of $(n_0, n_1)=(2500, 2500)$ (values provided in Table S1). The estimating procedure always provided coverages around the expected 95\% for the {\it sample-system} AUCs. There are excessively large coverage for Model II-a, where the underlaying AUC is close to 1. But, in general, we can say that the proposed procedure works adequately. The coverages of the real-system AUCs strongly depend on the underlying model. Not surprisingly, these coverages are around the 95\% in Model I-a, and Model IV-a (remember, real AUCs of 0.5). The distance between the sample-system and the real-system was also small in Model II, where we see also good coverage. However, they were poor in the rest of the models. Notice that, for instance, in Model III-(b, c, d), the real-system reported AUCs of 0.76, 0.79, 0.77, while the average AUCs obtained for the sample-systems ($(n_0,n_1)=$(200,100)) were 0.67 (ranging 0.40 to 0.86), 0.77 (ranging 0.26 to 0.88), and 0.67 (ranging 0.40 to 0.86), respectively. 

\begin{footnotesize}
\begin{table}
\caption{{\footnotesize {\bf Coverages.} Coverage percentages for the 95\% confidence intervals constructed based on the training-testing approach for the sample-system and the real-system (based on the Table S1 values) AUCs. We also report the average length of the obtained intervals. } \label{coverage}}

\begin{center}
\begin{footnotesize}
\hspace{-2cm}
\begin{tabular}{lcccccccccccccccccccccccc}
\rowcolor{blue!15}                 
                            &    \multicolumn{4}{c}{\bf{\emph{sample-system}}}                  &&  \multicolumn{4}{c}{\bf{\emph {real-system}}} && \multicolumn{4}{c}{\bf{\emph{ Length (mean)}}} \\   
\rowcolor{blue!15} & {\bf -a} & {\bf -b} & {\bf -c} & {\bf -d} && {\bf -a} & {\bf -b} & {\bf -c} & {\bf -d} && {\bf -a} & {\bf -b} & {\bf -c} & {\bf -d} \\   
{\bf Model I}        & \\
\rowcolor{blue!3}
\phantom{a} {\bf (50,50)}     &  95.6 & 94.0  & 91.0 & 93.6 && 95.4 & 59.2 & 62.0 & 62.5 && 0.274 & 0.135 & 0.160 & 0.132\\
\phantom{a} {\bf (50,100)}   &  93.6 & 92.5  & 92.7 & 90.4 && 93.3 & 57.8 & 54.1 & 55.9  && 0.236 & 0.102 & 0.126 & 0.100\\
\rowcolor{blue!3}
\phantom{a} {\bf (100,100)} &  93.8 & 92.0  & 92.1  & 90.9 && 95.2 & 56.5 & 36.8 & 61.2  && 0.193 & 0.077 & 0.103 & 0.074\\
\phantom{a} {\bf (200,100)} &  94.8 & 90.1  &  93.2  & 91.6 && 95.4 & 60.4 & 24.7 & 65.0  && 0.166 & 0.058 & 0.092  & 0.057\\
\rowcolor{blue!3}
{\bf Model II}       &&&&&&&&&&&&&&  \\
\phantom{a} {\bf (50,50)}    & 99.9 & 99.6  & 97.5 & 99.9 &&  93.7 & 96.9  & 88.5 & 94.8 && 0.043 & 0.040 & 0.081 & 0.045\\
\rowcolor{blue!3}
\phantom{a} {\bf (50,100)}   & 99.9 & 99.7 & 95.6 & 99.4 && 97.0 & 96.8 & 81.7 & 97.6 && 0.025 & 0.030 & 0.063 & 0.029\\
\phantom{a} {\bf (100,100)} & 99.9 & 99.3 & 95.0 & 100   && 99.1 & 96.6 & 87.9 & 99.3 && 0.016 & 0.027 &  0.046 & 0.021\\
\rowcolor{blue!3}
\phantom{a} {\bf (200,100)} & 100  & 99.2 & 91.8 & 99.6 && 100 & 99.5 & 93.4 &  99.5 && 0.010 & 0.014 & 0.034 &  0.012\\
{\bf Model III}      & \\
\rowcolor{blue!3}
\phantom{a} {\bf (50,50)}     & 99.6 & 94.0 & 92.3 & 93.6 && 99.6 & 63.3 & 71.4 & 56.2 && 0.058 & 0.263 & 0.242 & 0.263\\
\phantom{a} {\bf (50,100)}   & 95.4 & 92.6 & 94.5  & 93.7 && 99.3 &  58.1 & 80.1 & 55.6  && 0.048 & 0.222 & 0.202 &  0.221\\
\rowcolor{blue!3}
\phantom{a} {\bf (100,100)} & 93.1 & 94..2 & 94.3 & 94.2 && 97.4 & 53.7 & 84.9 & 49.1 && 0.041 &  0.180 & 0.159 & 0.179  \\
\phantom{a} {\bf (200,100)} & 91.1 & 94.0 & 94.7 & 95.6 && 92.5 & 42.6 & 88.7 &  40.6 && 0.035 & 0.158 & 0.139 & 0.158   \\
\rowcolor{blue!3}
{\bf Model IV}      &&&&&&&&&&&&&&  \\
\phantom{a} {\bf (50,50)}     & 95.5 & 92.6 & 93.4 &94.5  && 95.5 & 67.5 & 71.4 & 66.6 && 0.283 & 0.176 & 0.199 & 0.196\\
\rowcolor{blue!3}
\phantom{a} {\bf (50,100)}   &  94.8 &  92.7 & 92.9 & 92.0  && 95.1 & 66.7  & 72.4  & 60.0 && 0.243 & 0.162 & 0.160  & 0.159   \\
\phantom{a} {\bf (100,100)} &  94.7 &  94.1 & 93.6 &  94.7  && 93.8 & 54.1 & 64.1 & 55.8 && 0.197 & 0.129 & 0.130 & 0.128 \\
\rowcolor{blue!3}
\phantom{a} {\bf (200,100)} &  93.7 & 94.5 & 93.2 & 95.2 && 94.1 & 46.9 & 63.4 & 46.5 && 0.170 & 0.109 & 0.110 & 0.108  \\
\end{tabular}
\hspace{-2cm}
\end{footnotesize}
\end{center}
\end{table}
\end{footnotesize}

\section{Real-world application}
\begin{figure}
\centerline{\includegraphics[width=15.5cm]{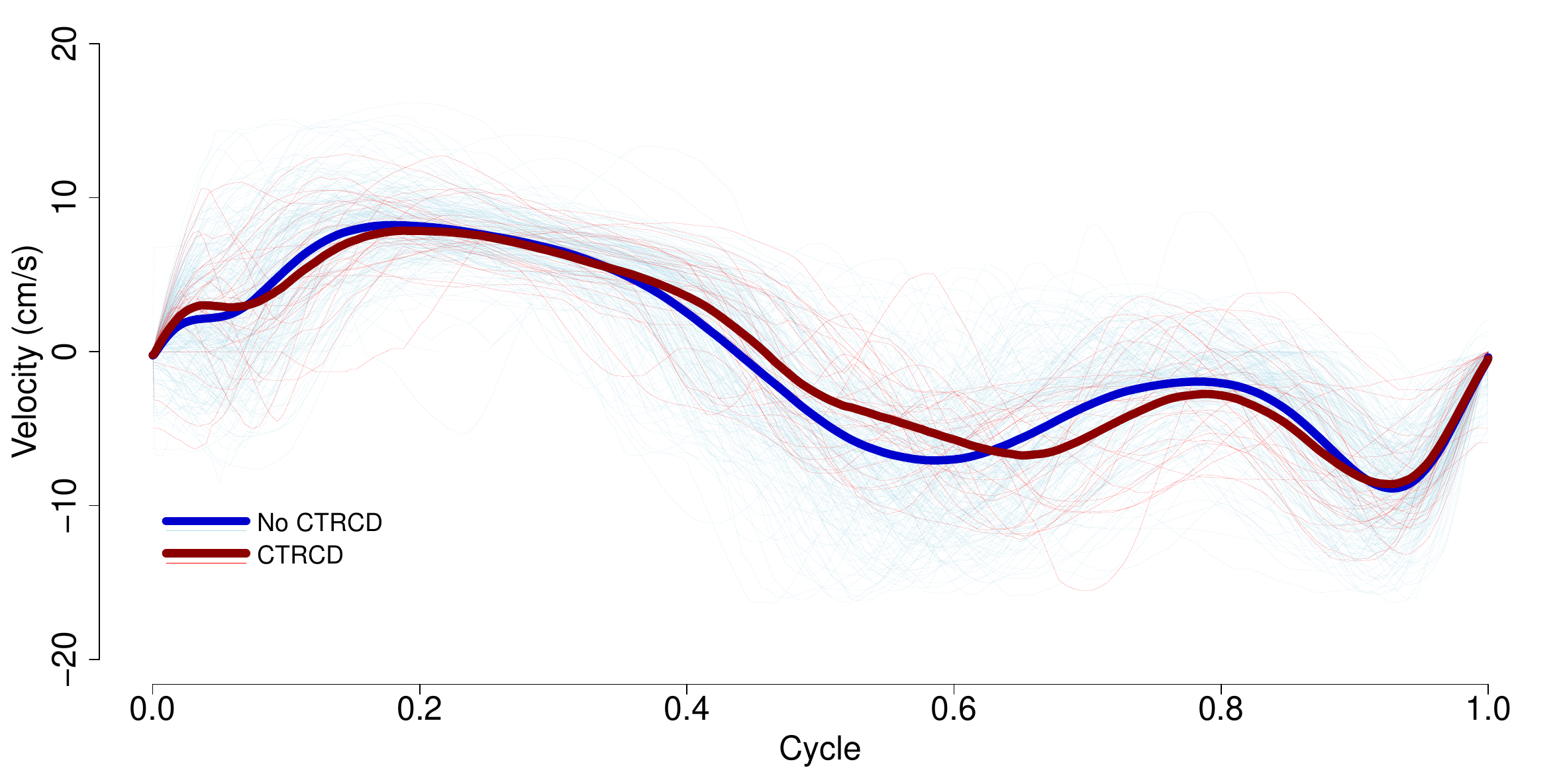}}
\caption {{\bf CTRCD data.} Average behavior (thick lines) of velocity function (VMM) based on the 243 patients without CTRCD (blue line), and on the 27 patients with CTRCD (red line). Thin lines represent each single patient trajectory.\label{example}}
\end{figure}

We are interested in studying the ability of the {\it velocity of myocardial motion} (VMM) for diagnosing cancer therapy-related cardiac dysfunction (CTRCD) in breast cancer patients. Therapies based on drugs targeting the HER2 protein have shown benefits for specific types of breast cancer in terms of overall response of patients and survival expectancy. However, associated cardiotoxicity (CTRCD) can be a major adverse effect. With the goal of understanding the ability of VMM for diagnosing CTRCD, as we anticipated in Section 1, we consider the data presented in Pi\~neiro-Lamas et al. \cite{pineiro23} containing measured of VMM using the echocardiographic technique called Tissue Doppler Imaging (TDI), which shows velocity as a function of time. Figure \ref{example} shows the average behavior (thick lines) of this velocity function based on the 243 patients without CTRCD (blue line), and on the 27 patients with CTRCD (red line). Thin lines represent each single patient trajectory. Average behaviors look similar between the groups, while the non CTRCD patients trajectories have more variability.

\phantom b\par

The AUCs for the Min., Max., and Int. were 0.68 (95\% CI 0.58 to 0.78), 0.55 (0.42 to 0.67) and 0.53 (0.41 to 0.65), respectively. When we apply the Lin., and Qua. criteria, the AUCs were 0.65 (0.55 to 0.75), and 0.72 (0.65 to 0.79) respectively. Remark: the last two models are not protected for overfitting (see Supplementary Material for an additional Monte Carlo simulation scenario including the particularities of this dataset). When we compute the PBC estimator using the whole dataset for both training and testing, we have an AUC of 0.60 (0.54 to 0.66), which would be potentially overfitted as well. The standard deviation involved in the confidence intervals for Lin., Qua., and PBC were computed using 200 Bootstrap iterations. AUCs provided for $\Upsilon_{\text{LIN}}$, and $\Upsilon_{\text{QUAD}}$ in \citet{bianco24} were 0.71 and 0.89, respectively.

\begin{figure}
\begin{center}
\begin{tabular}{cc}
\includegraphics[height=7cm]{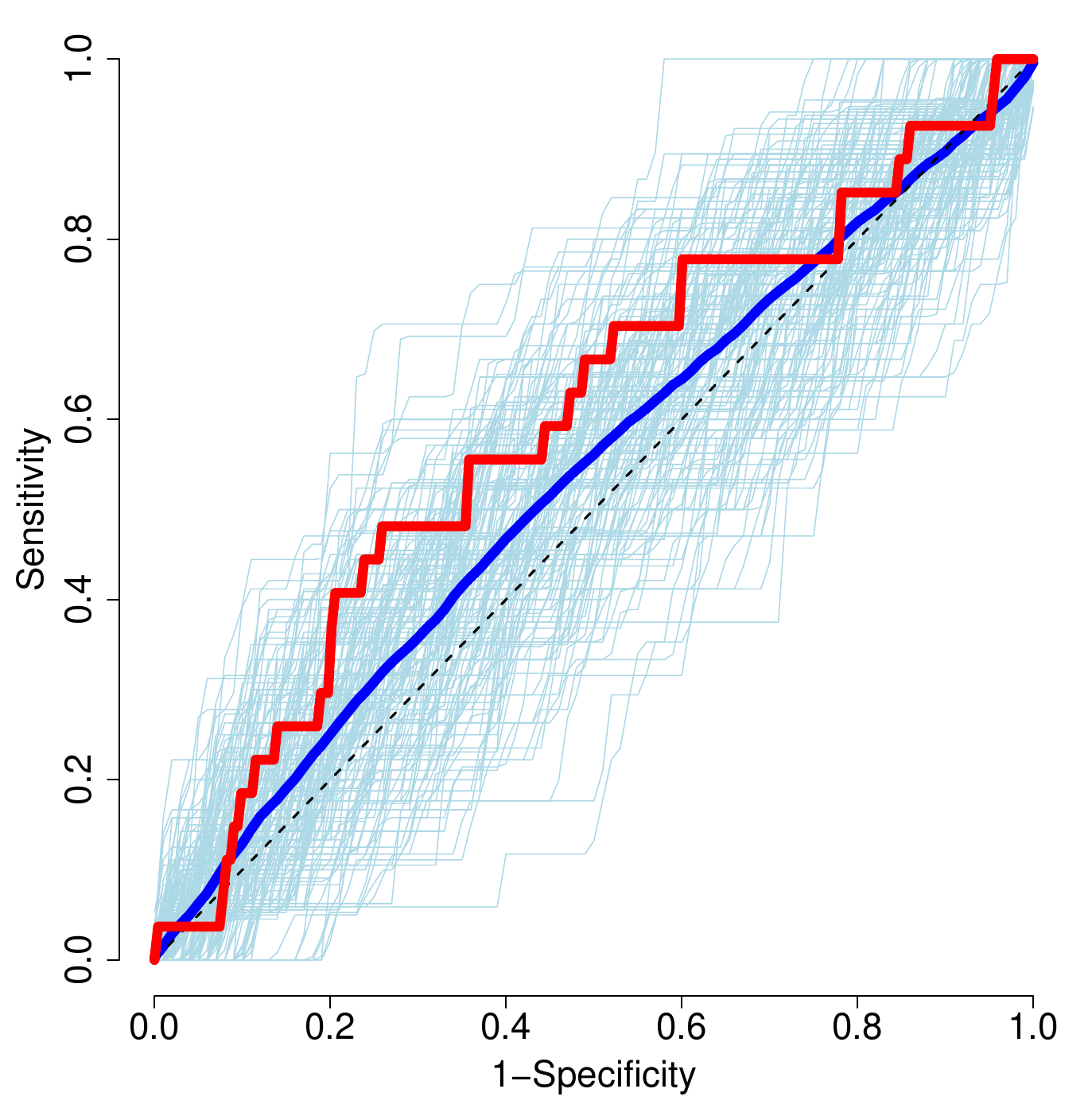} & \includegraphics[height=7cm]{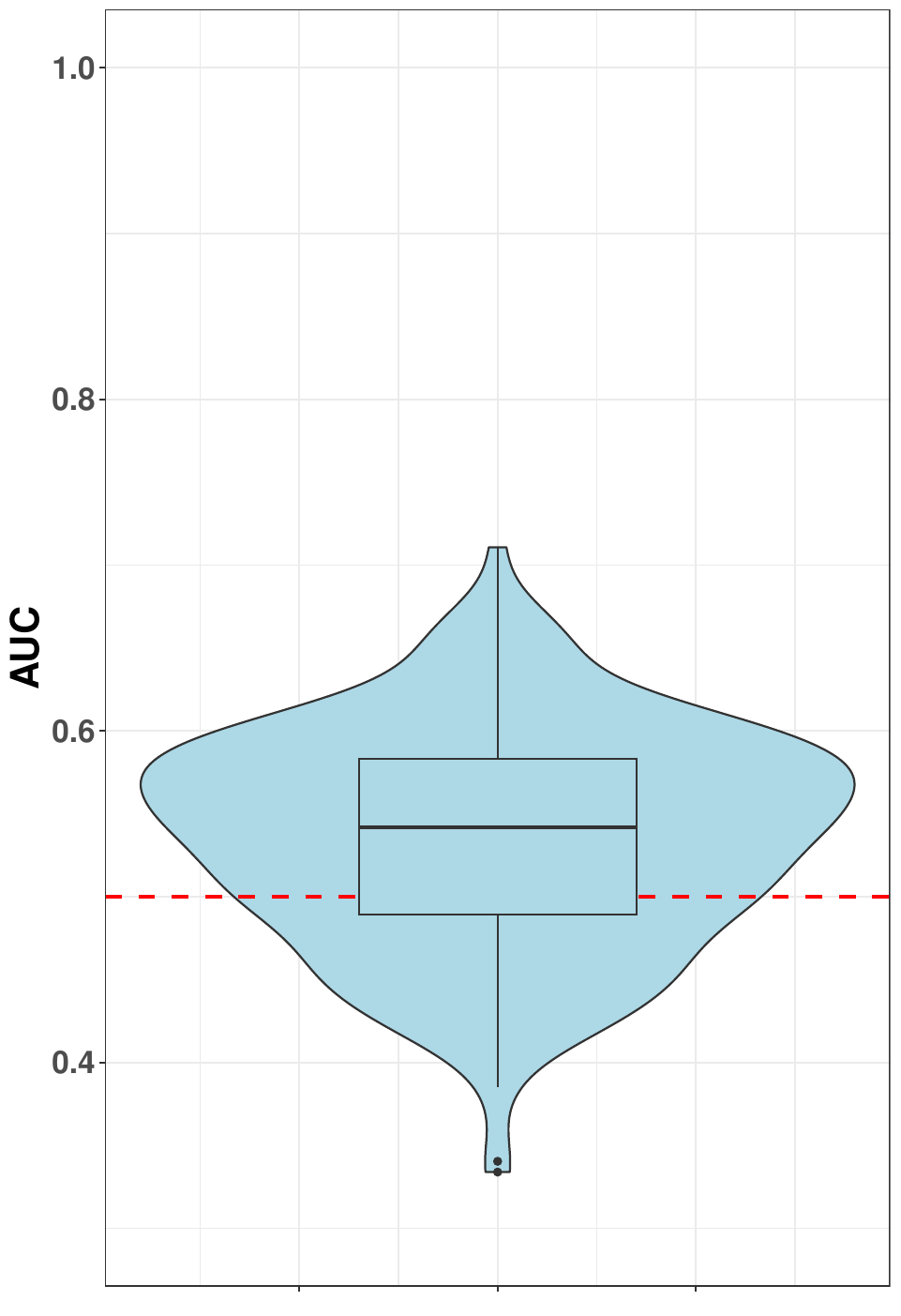}
\end{tabular}
\caption {{\bf Training-testing.} (Left) Average (blue-thick), and individual (thin) ROC curves for the 200 training-testing iterations (left), red-think line represents the ROC curves for the PBC procedure on the whole sample. (Right) Violin plot for the obtained AUCs (right).\label{training}}
\end{center}
\end{figure}

Now, we implement the explored training-testing approach. We randomly select 1/3 of the sample for training (approximately 80 negative and 9 positive curves), and check the quality of the resulting {\it sample-system} for identifying CTRCD patients on the remaining 2/3 of the sample. We repeat the process 200 times. The average AUC was 0.54 (95\% CI) (0.42 to 0.66, 95\% CI computed as an average of the lower, and upper bounds of the 200 95\% CI), ranging from 0.33 to 0.71. Figure \ref{training}-right shows the average (thick line) and the 200 ROC curves rays (thin lines). At left, we observe the violin plot for the AUC distribution.
\phantom b\par

\begin{figure}
\centerline{\includegraphics[width=14.5cm]{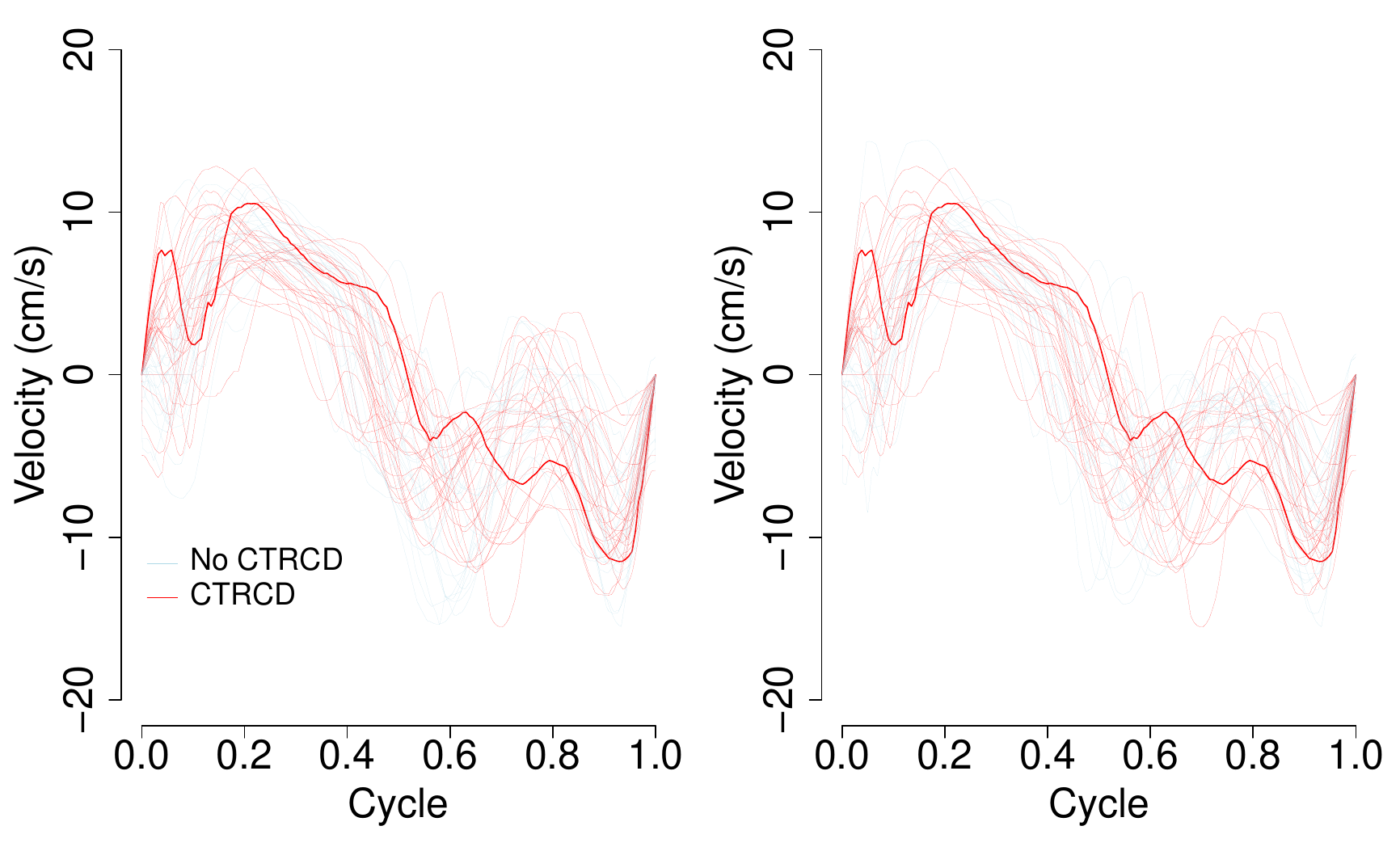}}
\caption {{\bf CTRCD data.} Trajectories from the 27 patients with CTRCD (red lines), and two 27 rays randomly selected from patients without CTRCD (blue lines).\label{exampleB}}
\end{figure}

Therefore, the actual conclusion regarding this data could be that, {\it (1)} based on the reported AUC, the PBC criterion does not get an adequate use of the information contained in the VMM rays, and in this case, with the available data, using other criteria would be better, or {\it (2)} the behavior of the VMM rays does not allow to accurately identify CTRCD and, the high AUCs observed (provided by Qua., $\Upsilon_{\text{LIN}}$, and $\Upsilon_{\text{QUAD}}$) are spurious findings caused by overfitted models. In order to figuring out the real differences between the involved trajectories, Figure \ref{exampleB} show the 27 rays from the patients with CTRCD (red lines), and two sets of 27 rays randomly selected from the patients without CTRCD. 

\phantom b\par

\section{Conclusions}
The presence of functional data (FD) is becoming more and more frequent in all scientific research fields and, particularly, in biomedicine. The adequate use of the resulting patient-trajectories with different goals represents a methodological challenge. In this paper, we are interested in applying FD to the binary classification problem (BCP). Of course, we are not the first considering this problem; \citet{escabias07}, \citet{estevez21}, \citet{jang23} or \citet{bianco24}, among others, have provided valuable inside on this topic. The crucial step for adapting the already existing techniques for the BCP to FD is to stablish an order on the obtained trajectories. Here, we propose to use the $L_2$-distance for approximating the probability that one function comes from the positive population (against of coming from the negative population), and use this probability for ranking the trajectories. The use of the area under the gROC curve, gAUC, allows to correctly classify trajectories in clustered situations. That is, for classifying a trajectory as positive/negative, we do not need that this is close to all the positive/negative trajectories, or to an average trajectory, rather to a group of them. The proposed non-parametric estimation procedure involves the selection of different parameters, which can lead to a considerable amount of overfitting. In order to control this, we implement a standard machine learning training-testing strategy, in which we use a portion of the sample for learning the parameters required for being used in the remaining portion.

Our Monte Carlo simulation study shows that the training-testing approach effectively controlled the potential overfitting of the PBC criterion (Model I-a, and Model IV-a). Besides, PBC was the clear winner in the considered Model II and Model IV. It seems that our proposal behaves well in situations where there exists more than one type of trajectory for group. The current approach, taking only one random selection for the model construction, made the variability very large; even in scenarios with an underlying AUC close to 1, and with a very accurate overall behavior, we can see very low AUCs in some of the iterations. The percentage of outliers reduces when the sample size increases. This problem could be mitigated generating more random sample-systems in a more machine-learning base (see, in Section 5, the real-world problem implementation).

In the analysis of the considered real-world problem, the PBC criterion does not reach very good results. The overall AUC is lower than those obtained by procedures such as Qua., and 
$\Upsilon_{\text{QUAD}}$, and similar to $\Upsilon_{\text{LIN}}$; the training testing approach reported an average AUC similar to Min., and Lin. We did not consider $\Upsilon_{\text{LIN}}$, and $\Upsilon_{\text{QUAD}}$ in our Monte Carlo simulations (software not available), but we have seen the overfitting of Lin., and (specially) Qua. in the simulation process (see the online Supplementary Material for a scenario informed by the real-world example). Besides, Figure \ref{exampleB} shows how, when the involved sample sizes in the positive and the negative groups were equal, the overall visual impression about the differences between the rays changes and, although some of the negative curves present lower values first (about cycle 0.45-0.6) than the positive rays (around cycle 0.55-0.66), and the lowest VMM value on the negative uses to be lower than the lowest value on the positive (the AUC of the Min. criterion was 0.68), this is not the case for a number of these trajectories. Based on the raw of ROC curves (Figure \ref{training}-right), in this problem we have negative curves closer to the positive than to the negative populations.

In summary, we think the current paper presents a clear and reasonable solution for the use of functional data in the binary classification problem. The criterion can be easily extended/adapted to other contexts, and could be easily included in a learning process in which the {\it sample-systems} are storage and ready to be used for informing decisions. As usual, the employed criterion is not universally superior to others but allows a clear understanding of the classification process and makes a correct use of the available information. Unfortunately, the R code we developed is excessively time-consuming (partially, because the use of the gROC curve in an overall setting) and, probably, more skilled programmer should take care of its technical implementations.

\section*{Appendix}
\noindent {\it Theorem's 1 proof.}
First, we are going to prove that the eCDF derived from $\hat{\cal P}_n$ consistently estimated CDFs of $\cal P$ in the negative and in the positive populations. From Mart\'inez-Camblor et al. \cite{camblor22}, we know that, for each $1\leq i\leq n$, $\hat p_{f_i,n}= p_{f_i} - \epsilon_{i,n}$, where $\sqrt{n}\cdot\epsilon_{i,n}$ is an asymptotically normal distributed random variable with variance $\sigma^2_{i,\lambda}<\infty$. Therefore, for $1\leq p\leq 1$
\begin{small}
\begin{align}
\frac{1}{n}\sum_{j=1}^{n}{\mathds 1}\{\hat p_{f_i,n}<p\} - F(p) =& \frac{1}{n}\sum_{j=1}^{n}{\mathds 1}\{p_{f_i}<p + \epsilon_{i,n}\} - F(p) \nonumber \\
\leq & \frac{1}{n}\sum_{j=1}^{n}{\mathds 1}\{p_{f_i}<p + \epsilon_{M,n}\} - F(p + \epsilon_{M,n}) + F(p + \epsilon_{M,n}) - F(p) \nonumber \\
\leq & \left (\frac{1}{n}\sum_{j=1}^{n}{\mathds 1}\{p_{f_i}<p + \epsilon_{M,n}\} - F(p + \epsilon_{M,n})\right)+ \left( F(p + \epsilon_{M,n}) - F(p) \right), \label{ineq1}
\end{align}
\end{small}
where $\epsilon_{M,n}=\max_{0\leq i\leq n}\{\epsilon_{i,n}\}$. Analogously, if $\epsilon_{m,n}=\min_{1\leq i\leq n}\{\epsilon_{i,n}\}$
\begin{small}
\begin{align}
\frac{1}{n}\sum_{j=1}^{n}{\mathds 1}\{\hat p_{f_i,n}<p\} - F(p) \geq & \left (\frac{1}{n}\sum_{j=1}^{n}{\mathds 1}\{p_{f_i}<p + \epsilon_{m,n}\} - F(p + \epsilon_{m,n})\right)+ \left( F(p + \epsilon_{m,n}) - F(p) \right).\label{ineq2}
\end{align}
\end{small}
Inequalities (\ref{ineq1}) and (\ref{ineq2}), the Glivenko-Canteli Theorem, and the continuity of $F(\cdot)$ guarantee that
\begin{equation*}
\sup_{p\in\mathbb R}|\hat F_{n,n}(p) - F(p)|\longrightarrow_n\, 0\quad a.s.
\end{equation*}
Therefore, we have that $\sup_{p\in\mathbb R}|\hat F_{n_0,n}(p) - F_0(p)|\rightarrow_{n_0}\, 0\, (a.s.)$, and $\sup_{p\in\mathbb R}|\hat F_{n_1,n}(p) - F(p)|\rightarrow_{n_1}\, 0\, (a.s.)$. For the consistency of the ROC curve estimator, we will argue as in D\'iaz-Coto et al. \cite{susana21}. First, we know that for each sequence $\{u^*_N\}_{N\in\mathbb N}\in (0,1)$, it holds the equality
\begin{equation}
[{\cal R}(p_N^*] - {\cal R}(p)] = r(p)\cdot [p^*_N - p] + {\cal O}(p^*_N - p),
\end{equation}
where $r(\cdot)$ is the first derivative of ${\cal R}(\cdot)$. Let be $q^*_{n_0}=\hat F^{-1}_{n_0,n}(1-p)$, and $p^*_{n_0}= 1-F_0(q_{n_0}^*)$, then
\begin{align*}
[\hat {\cal R}_{n,\bullet}(p) - {\cal R}(p)] = & [\hat {\cal R}_{n,\bullet}(p) - {\cal R}(p^*_{n_0})] + [{\cal R}(p^*_{n_0}) - {\cal R}(p)]\nonumber\\
=& [F_1(q^*_{n_0}) - \hat F_{n_1,n}(q^*_{n_0})] + r(p)\cdot [\hat F_{n_0,n}(q^*_{n_0}) - F_0(q^*_{n_0})] + {\cal O}(p^*_{n_0} - p),
\end{align*}
and the proof is immediate from the uniform convergence of the involved eCDF. \hfill $\Box$\par
\phantom b\par\noindent
Proofs of Theorem 2, and Corollary 1 can be directly derived from \citet{hsieh96}.

\section*{Online supplements}
As online supplementary material, we provide the \url{R} code used in Section 4, and some complements to our Monte Carlo simulation study.

\section*{Data Availability Statement}
Data used are freely available. Information required for accessing the data are provided in the manuscript.

\section*{Conflict of Interest}
The author does not have conflicts of interest to report.

\section*{Funding} \label{Sec:Funding}
PM-C disclosed receipt of the following financial support for the research, authorship, and/or publication of this article: This work
was supported from the Grant PID2023-148811NB-I00 from Agencia Estatal de Investigaci\'on (Ministerio de Ciencia, Innovaci\'on y Universidades, Spanish Government).

\section*{ORCID iD}
Pablo Mart\'inez-Camblor \url{https://orcid.org/0000-0001-7845-3905}.

\bibliographystyle{unsrtnat}
\bibliography{Bibliography}
\end{document}